# A new perspective on the prediction of the innovation performance: A data-driven methodology to identify innovation indicators through a comparative study of Boston's neighborhoods

Eleni Oikonomaki[1], Dimitris Belivanis[2]

[1] URENIO Research, Aristotle University of Thessaloniki, 54124 Thessaloniki, Greece
Northeastern Boston University, 360 Huntington Ave, Boston, MA 02115
elenoikonomaki@gmail.com (E.O.);

[2] Dbelivanis@stanford.edu; (D.B.)

## Abstract

In an era of knowledge-based economy, commercialized research and globalized competition for talent, the creation of innovation ecosystems and innovation networks is at the forefront of efforts of cities. In this context, public authorities, private organizations, and academics respond to the question of the most promising indicators that can predict innovation with various innovation scoreboards. The current paper aims at increasing the understanding of the existing indicators and complementing the various innovation assessment toolkits, using large datasets from non-traditional sources. The success of both top down implemented innovation districts and community-level innovation ecosystems is complex and has not been well examined. Yet, limited data shed light on the association between indicators and innovation performance at the neighborhood level. For this purpose, the city of Boston has been selected as a case study to reveal the importance of its neighborhood's different characteristics in achieving high innovation performance. The study uses a large geographically distributed dataset across Boston's 35 zip code areas, which contains various business, entrepreneurial-specific, socio-economic data and other types of data that can reveal contextual urban dimensions. Furthermore, in order to express the innovation performance of the zip code areas, new metrics are proposed connected to innovation locations. The outcomes of this analysis aim to introduce a 'Neighborhood Innovation Index' that will generate new planning models for higher innovation performance, which can be easily applied in other cases. By publishing this large-scale dataset of urban informatics, the goal is to contribute to the innovation discourse and enable a new theoretical framework that identifies the linkages among cities' socio-economic characteristics and innovation performance.

## 1. Introduction

Over the last decades, innovation has increased its importance within the pattern of economic growth, moving to the central stage of economists and policymakers concerning the factors that enable the process. According to the "innovation-based growth theory", economic prosperity results from increase in knowledge, scientific and technological improvements, along with the development of an effective private-public partnership. The origin of the innovation-based growth models goes back to Romer [1,2,3] where growth is driven by specialization and increasing division of labor. Innovation performance is therefore considered a crucial determinant of competitiveness and national progress in the 21st century globalized economy [4]. Cities and innovation are nowadays strongly linked. The emerging trend of innovation districts globally is another indication of cities' efforts to become a favorable globalized environment for innovation to prosper.

Although there are research studies attempting to examine the dynamics that lie behind the creation of an innovation district, less emphasis has been given on the synergies and socio-economic characteristics of the actors coexisting in the district or being involved in its development process. The existing empirical studies are qualitative case studies, such as those by Spigel in Canada [5] and Mack and Mayer in the US [6] and studies measuring innovation ecosystems with quantitative data, such as the study by Ács et al., 2014; and Radosevic and Yoruk, 2013 [7,8] and several others mostly focusing on national and regional level [9, 10].

This research area still lacks empirical evidence at a smaller than regional scale, and more specifically at the functional region, neighborhood level. Several scholars support that there is substantial variation between different localities of the same region [11, 12]. Local innovation indexes are considered to be useful sources of information for policymakers, given the fact that microregions appear to have specific strengths and weaknesses in terms of innovation and therefore they face specific problems and require differentiated policies even within a given region [13,14].

Autant-Bernard & LeSage have also identified that innovation performances change dramatically between different areas within a region [15]. In the same direction, Porter supports the idea that a country's innovative capacity depends on the more specific innovation environments present in a country's industrial clusters, whether firms invest and compete on the basis of new-to-the-world innovation depends on the microeconomic environment in which they compete, which will vary in different fields [16].

**1.2 Research Goals**

The current paper discusses literature on innovation assessment in order to define the research question context and the innovation strategies cities put into action to become ecosystems for innovation and entrepreneurship. It builds upon previous efforts to identify new ways and more efficient technological tools to understand the dynamics of neighborhoods with high performance in terms of entrepreneurship and innovation [17]. A previous study suggested that the existing metrics-based toolkits focus mainly on innovation indexes from a top-down perspective and hardly measure innovation at a smaller than the regional scale, almost neglecting the fact that innovation is mainly built from bottom-up processes. This study focuses on innovation success factors at the city/neighborhood level, given the fact that this is generally seen as the more adequate level of analysis in terms of proposing policy [18, 19] and nurturing entrepreneurship [20, 21, 22].

This research aims at evaluating the dynamics of districts/neighborhoods with high activity of startups and entrepreneurship, using big data analytics strategy as our main method. Ultimately resulting in a critical explanation of our own proposals for detecting innovation success factors. In support of this effort, data has been collected from several open-source web mapping and other types of platforms for entrepreneurship, demographics, socio-economic data.

Our analysis is divided into three parts corresponding to three scales of inquiry. Initially we provide a short literature review on innovation success factors and more specific initiatives implemented by the city of Boston towards this direction. Then Boston's neighborhoods were looked at zip-code level in regard to their business activity. We elaborate on the creation of a dataset consisting of the business projects permits, zip codes, project description, type of occupancy, the census id of each business permit, merged with other socio-economic data published by the census bureau and other Boston GIS files. We propose a model that captures key factors that hold back or enhance innovation performance across the different neighborhoods. After explaining the method, we check the results and generate a few tables and maps emphasizing Boston's neighborhood characteristics.

Finally, in the findings section, we briefly explain the outcome of this paper, to respond to our research question. The immediate locale of each neighborhood was explored to weigh the advantages and disadvantages of their respective environments for entrepreneurs and innovation activity, but also the socio-economic characteristics, in a way to better understand the high concentration of innovation hubs, incubators, accelerators at specific areas of the city. By highlighting the different neighborhoods' distinctive socio-economic features, strengths, and shortcomings, we aim at providing a conceptual discussion of the notion of 'Neighborhood Innovation Index' and elaborate why and how high city and neighborhood-level entrepreneurship shows systemic characteristics.

More specifically, the paper sets out the findings of the study as follows: Section 1 introduces the topic of the existing innovation indicators and the level of innovation assessment and identifies the gap for empirical evidence at the local/neighborhood level. (Section 1.1); The section also describes the main research goals and proposes a theoretical framework that will provide support to policy makers to improve the current tools they use to better understand the success of innovation districts (Section 1.2).

Section 2 analyzes the indicators already discussed in literature (Section 2.1), which are considered important ingredients for successful innovation districts. It also contains an analysis of Boston's neighborhoods as local clusters of innovation (Section 2.2) and discusses previous studies assessing Boston's neighborhoods in terms of their entrepreneurial and socio-economic characteristics (Section 2.3).

Section 3 attempts at introducing a new method for analyzing innovation determinants by incorporating publicly available datasets with the creation of a customized dataset concerning innovation. For this purpose, information was scrapped from the web and queried from different apis. Those extracted features are used as innovation indicators and their connection with the available socioeconomic data is examined at the zip-code level. It also contains an overview of all open-source datasets and features used for this analysis (Section 3.1) and visualizations of all the study's key findings that indicate feature collinearities and neighborhood patterns (Section 3.2).

Section 4 discusses the main constraints we faced during our research presented in Chapter 3. It concludes with recommendations for improving our method for the prediction of innovation determinants.

## 2. Success factors of innovation ecosystems: Literature and Theory Framework

Having explored some of the broader issues associated with the need for metrics that can capture innovation performance at the neighborhood level, we will now take a closer look at innovation determinants that are mentioned in bibliography.

The Brookings Institution provides a commentary on the urban model that is now emerging, giving rise to what is called "innovation district." As described "these districts, by definition, are geographic areas where leading-edge anchor institutions and companies cluster and connect with start-ups, entrepreneurs, business incubators and accelerators. Districts are also physically compact, transit-accessible, and technically wired and offer mixed-use housing, office, and retail". Buildings typically located at the heart of the community, become inclusive centers for the local entrepreneurs. New housing developments for mixed income and skill training centers training residents on technologies/skills that match district region employment, are also contributing to the success of the district.

Innovation ecosystems are significantly dependent on talent attraction and retaining, on an entrepreneurial and risk-taking culture, as well as on the presence of R&I infrastructure and on compatible and complementary system stakeholders. The orchestrators or main key-actors play an essential role in the innovation ecosystems, influencing directly and indirectly their development. The Quadruple Helix Model of innovation, a model originally conceptualized by Elias Carayannis and David Campbell, recognizes four major actors in the innovation system: science, policy, industry, and society. On the one side, the anchor Higher Education (HE) institutions that nurture the Entrepreneurial Environment, an entrepreneurial approach emerging from the local governments, eliciting risk taking and bottom up civic participation in tackling key issues in the city, a fully operational networking structure of some intermediary actors supporting entrepreneurial collaboration, cross-fertilization and co-creation, which collaborate at multiple- scales; and local authorities supporting urban regeneration initiatives complementing economic development initiatives and making available civic-led spaces enabling grassroot collaboration and cooperation for the society [24].

Local, regional, national, and international innovation-related policy agendas have a relevant impact on the strategic directions of innovation ecosystems, such as the UN 2030 Agenda for Sustainable Development. Internationalization is another core element for competitive, sustainable, and successful strategies of local and regional innovation ecosystems [23].

### 2.1 Innovative Neighborhoods of Boston: socio-economic and entrepreneurial neighborhood characteristics

The Greater Boston area is currently one of the most innovative locations in the US local development landscape, thanks to its high agglomeration of educational institutions and industries. The entire urban region, which is recording the highest rate of growth anywhere in the US [25], is increasingly able to attract the interest of major investors. Over the last three decades, the cities of Boston and Cambridge, public and private investments have prioritized the boost of sectors such as education, financial services, life sciences, and the high-tech industries, which today represent the main clusters within the entire urban region. Greater Boston is a huge metropolitan area that includes the counties of Suffolk and Middlesex. The municipalities of Boston and Cambridge are located within it. In 2016, the Greater Boston area was the foremost location in the U.S. for fostering entrepreneurial growth and innovation [26]. This ranking is based on how well the top 25 US metropolitan areas "attract talent, increase investments, develop specializations, create density, connect the community and build a culture of innovation" [26].

Even though the San Francisco Bay Area dominates most index categories, "its lack of a collaborative community and a declining quality of life for wide swaths of its citizens vaulted Boston to the top spot, ". The survey also ranks the Bay area "quite low in terms of quality of life (22nd), which may be reflective of the increasing cost of living". At the same time, Boston ranked second to the Bay Area "on most traditional metrics of start-up activity, but local entrepreneurs indicate stronger connections with universities, institutions and citizens" [26]. The cities of Boston and Cambridge host significant Higher Education institutions' start-ups, tech industries and research centers which make them the ideal innovation urban ecosystem attracting high-skilled creative workers, innovators, and investors. Moreover, the local government contributes to the growth of the innovation ecosystem by providing services, funding and resources for businesses and entrepreneurs (e.g., the Cambridge Entrepreneurship Assistance Program, the Cambridge Small Business Enhancement Program), implementing planning initiatives and creating new spaces for the local innovation community.

There are several public-private partnerships involving collaboration between Boston city's government and private-sector that aim to enhance innovation. Financing a project through a public-private partnership can allow a project to be completed sooner or make it a possibility in the first place such as the District Hall, Roxbury Innovation Center. On the other hand, Cambridge Innovation Center, Masschallenge are considered successful private –led initiatives, while Kendall Square, Seaport District, Dudley Square have been initiated as public-driven regeneration initiatives.

**LifeTech Boston**
In 2004, the Boston Planning Development Agency (BPDA), then known as the Boston Redevelopment Authority (BRA), launched the 'LifeTech Boston' policy initiative.This was a significant incubatory step towards the eventual creation of a new redevelopment model, the first Innovation District in Boston. The most significant strand of the original strategy was later identified in the 'Boston Innovation District' (BID), a planning initiative launched in 2010 by the Menino administration and still in progress..

**Boston Innovation District**
The BID project aims to create a complex neighborhood that mimics the success of 22@Barcelona, with the ultimate goal to attract financers, resources, and talent. The BID project was conceived to redevelop the South Boston Waterfront, an underutilized area of 1,000 acres that previously hosted industrial activities and parking and transform it into a thriving hub of innovation and entrepreneurship together with new residential, commercial, and retail spaces (about 7.7 million sq. ft.) with a mixed-use configuration.

**District Hall**
District Hall opened in 2013 and was founded through a unique public-private partnership in response to the City's call. First project in their 23-acre master plan for Seaport Square. The city invited the Cambridge Innovation Center (CIC) to create an active innovation hub concept. CIC then developed the initial concept and design, as well as provided construction management and financial support for the project. CIC has ongoing responsibility and holds the lease for the property. The Venture Cafe ́ Foundation was involved in the initial planning process and continues to manage operations and programming since 2013.

**Cambridge Innovation Centers**
Cambridge Innovation Centre (CIC) and the Venture Cafe ́ Foundation were both founded by Tim Rowe, the former in 1999 as an incubator, the latter in 2010 as a social experiment. CIC is a private entrepreneurial activity based on renting shared and flexible office spaces with an innovative style. It currently hosts over 700 companies across two buildings, located in Kendall Square and in downtown Boston, about 500 of which are start-ups.

**Kendall Square**
Kendall Square is a former brownfield located in Cambridge (MA), opposite the Charles River. It started in 1868 as an industrial district and consolidated this function with the opening of the first underground line nearby. The presence of the Massachusetts Institute of Technology dates to 1916. Following the Second World War, the area entered an era of decline, which the Cambridge Redevelopment Authority (CRA), established in 1955, sought to reverse also through the clearance of 29 acres of land for the accommodation of NASA.

**Dudley square**
The 'Neighbourhood Innovation District' (NID) is an on-going public strategy launched in 2014 by the government of Boston City Council. The main goal of this initiative is to encourage and spread innovation and technology within low-income neighborhoods to improve small business growth and local economic development. Following specific criteria mentioned in the innovation district literature (transit access, affordable office space, arts and cultural amenities, involvement of non-profit organizations) and considering the features of the area (e.g., the presence of higher-education institutions, vacant lots, transportation nodes) the location for the first experiment was chosen to be the 'Dudley Square-Upham's Corner Corridor', a vibrant zone within the Roxbury neighborhood.

**Roxbury Innovation Center**
Roxbury Innovation Center (RIC) was created through a public-private partnership with the City of Boston and The Venture Café Foundation. Its mission was to support local economic development, in Roxbury, Dorchester and Mattapan by empowering and guiding innovation and entrepreneurship, as viable career options. Since opening in late 2015, RIC has provided a diverse variety of resources for small business owners of all stages and industries, through instructional workshops and courses, networking events, and office hour mentorship.

**MassChallenge**

Headquartered in the United States with locations in Boston, Israel, Mexico, Rhode Island, Switzerland, and Texas, MassChallenge strengthens the global innovation ecosystem by accelerating high-potential startups across all industries, from anywhere in the world for zero-equity taken. As MassChallenge's flagship location, MassChallenge Boston has brought together corporates, policy makers, and innovation leaders to support and inspire the next generation of innovators. Over the past eight years, it accelerated more than 1,000 startups from across the world – startups that have made patients healthier, our communities more connected, and our communities safer.

**2.3 Previous studies assessing innovation across Boston's neighborhoods**

The MAPS-LED project is another project in the same direction, which emphasizes innovation as a critical element for the further development of the cities. More specifically, the project proposes a cluster mapping strategy that tries to localize the economic clusters to understand better innovation at the neighborhood level. The report of this EU funded project published in 2019 marked different areas of Boston's greater area related to the respective cluster hosted. For this current study, we are considering Boston's innovative neighborhoods, and other standard neighborhoods, based on the MAPS-LED project.

TREnD project, an EU funded project tries to shed light on Boston's resilience strategies, resulting in a better and more inclusive transition process for the adoption of the green innovative technologies. The project "Navigating the green transition during the pandemic equitably: A new perspective on technological resilience among Boston Neighborhoods facing the shock" published in 2023, conceptualizes this transition process as an evolutionary path.

**3. Innovation indicators at the neighborhood level: Visualizing the Relationships among Features and Innovation Performance**

**3.1 Scope, Methods, and Data Collection**

The study aims at contributing to the current research efforts in innovation ecosystems by providing a dynamic model to demonstrate the correlations between the innovation performance and neighborhood characteristics. The project's methodological framework strongly relies on exploring a significant number of observations in the City of Boston to diagnose underlying correlations and connections of the socioeconomic data to innovation indicators and illustrate the interdependencies in entrepreneurial activity and social characteristics. The study introduces the "Neighborhood innovation Index" and the use of big data analytics to support the innovation performance measurement at this subregional (local) level. It also explores new ways such analyses may afford more opportunities than were previously examined by innovation assessment toolkits.

Thus, having this in context to answer the research questions stated earlier, we used the Boston Permits dataset published by Boston Area Research Initiative (BARI) to capture appropriate information about neighborhoods' tendency toward investing in business projects. The permit database is an excellent data source to capture the Boston changes during seven years. Benefiting the quantitative and georeferenced observations, it would be facilitating to get the pulse of the city according to the level of projects.

For the purpose of creating an innovation index, available data on the web was utilized. Specifically, the Google Maps api, Openstreetmap api, and indeed api, was used. For the google api, the keywords shown in table 2 were used and all the locations were gathered with their corresponding unique id, number of ratings, and average rating. The data is available for each individual keyword, however the concatenation of all the keywords is considered and presented as the most indicative. It is worth pointing out that each location that corresponds to a unique id was considered only once even though it could appear in multiple keyword searches. The number of locations is the most indicative as it is directly expressing the number of businesses that are relevant to innovation, but the total number of ratings and the mean rating provided additional insights, the average rating of the zip code was weighted by the number of ratings for each location.

In addition, the openstreetmap api was used to augment the dataset with locations of innovation. The use of keywords was not possible but rather the category of the building was used specifically the following tags were used to query locations in Boston: (*company=startup, office=coworking, office=research*). Unfortunately, the dataset was not covering the whole range of zip code areas, however as expected the area where those locations existed coincided with the areas which had the highest presence in the Google Map dataset. Even though the dataset is not as complete it is worth investigating as it is an open source that also provides the evolution over time and links the locations of innovations directly and not through a keyword that could be misleading.

Finally, data from the website www.indeed.com was collected, a site where job postings are available. Specifically, all jobs related to technology were collected and they were assigned to a specific zip code where the job was located. Unfortunately, most of the job postings were not including the exact address which were disregarded as there was no available zip code, the solution of geo-decoding was considered but has not been implemented yet. The dataset was not covering all the zip codes of the municipality of Boston, but the location with the most job offering coincided once again with the location where the highest innovation activity was shown from google maps.

**Summary of Variables**

The building permits include variables that can be grouped into three categories: business characteristics, all of which are original to the data; geographic information, which includes variables introduced by BARI to make the data more easily compatible with other data sources; and types of work, categorized by BARI based on the types of permits granted. For the purpose of this study only permits that were considered commercial or mixed use were considered.

The census data was provided in census tract and it was aggregated to zip code areas, and as for the data collected from the Google and Openstreetmap the exact location was available and therefore a spatial merging was performed with the polygons of Boston zip code areas. An overview of the data used is given in Table 1.

**Table 1.** Features and source used for supporting the dataset.

| Entrepreneurship - Innovation Activity | Social Data | Economic Data |
|---|---|---|
| Start-up locations (gmap) | Race (Census Bureau) | Median Income(Census Bureau) |
| Start-up locations (openstreetmap) | Vacancy (Census Bureau) | Median Household value (Census Bureau) |
| Start-up locations (crunchbase) | | Job openings (indeed) |
| Business Permits (BARI) | | |

**Table 2.** Keywords used for the google search.

| innovation hubs | clustering | innovation center | startups |
| innovation districts | open innovation | tech hub | technology park |
| incubator | accelerators | regional innovation | co-working space |

**3.3 Research Findings**

The purpose of this study is first to identify and quantify through an index or indices of the innovation performance of each area. This was achieved through the data gathering for locations with innovation impact, and it was performed as described above. The values collected from Google maps were considered to be the most appropriate as they provide similar results with other proposed values, they are more reliable for areas with low innovation presence, and can be reproduced easily for other areas in the USA or in a different country.

Specifically in figure 1 the correlation of three proposed innovation metrics with the socio-economic data is presented, those are namely the number of innovation locations, the number of average ratings, and the total number of ratings of all the locations in the zip code area. The first observation that can be made is that there are no high collinearities, this could rise from the heterogeneity of the studied areas as some are purely residential and some are business districts. Once the dataset is augmented with more areas it would be interesting to consider those two types of areas separately.

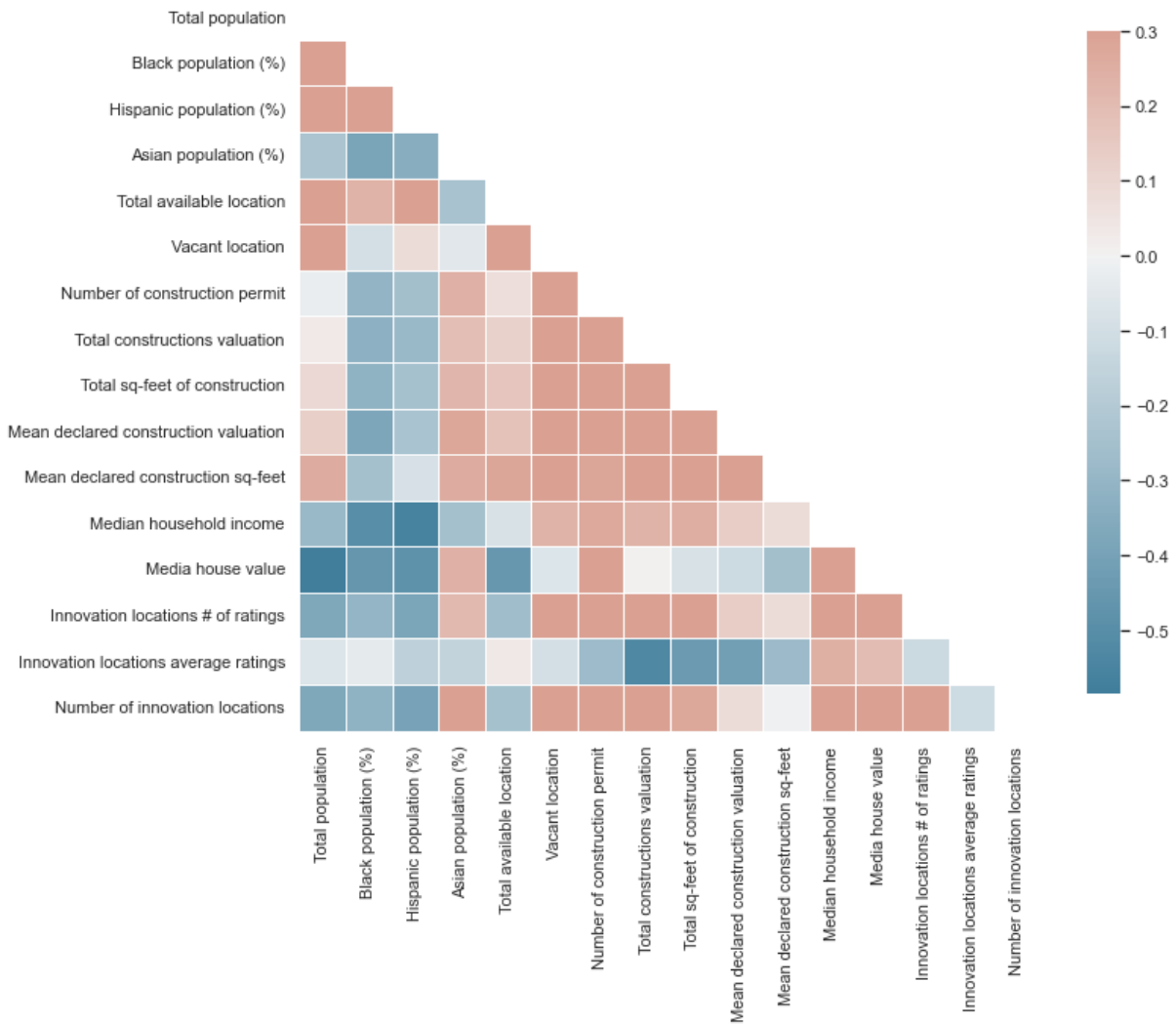

Figure 1. Matrix. Correlation coefficient is indicated by color,

Even though the correlations are relatively small (-0.5 - 0.3) the correlation matrix provides some interesting insights. There is a negative correlation of Black and hispanic population with innovation locations something that could be understood better when looking at figure 2, the locations that the number of innovation locations are high is in the city center where the business district is and it happens to have a higher percentage of white population as shown if figure 4, which can be explained historically and due to the elevated average price of the households in the central area. Another correlation that can be ascribed to the fact that most of the innovation locations are in business districts is its correlation with vacant spaces. Due to the volatility of the real estate market in business areas it makes sense that there are more vacant locations, something also seen in figure 3 where a map of vacant spaces is presented.

The number of constructions permits is positively correlated with the number of innovation locations, a bidirectional explanation can be provided as the more innovation locations and startup companies are attracted to an area the more construction permits and from the other side the areas with most business activity and therefore construction will attract more companies of which some will be related to innovation. An interesting finding that cannot be easily explained is the negative correlation of the average rating to the construction permit activity of an area, this should be studied further in more areas, as this could be an artifact from cases when the rating is not representative when the number of ratings is small for each location.

It is really interesting how the innovation indices are correlated among them. The first is that the rating is negatively correlated with the number of ratings, which could be explained as the rating is not representative if there are not enough ratings as mentioned above. The positive correlation of the number of locations and number of total reviews was anticipated as the number of locations increased the bigger the chance of review to be provided. However, the reason the correlation is low 0.3 and not almost collinear as someone could have assumed is because the number of innovation locations is linked to the average number of ratings per location therefore a super-linear relationship of number of locations and number of total reviews exists. This could be explained as collaboration among the innovators, because this leads to bigger visibility and presence in the local community that would possibly rate those locations. This implies that the areas where innovation is blooming will create a hub that would later attract more innovators and this becomes a positive feedback loop.

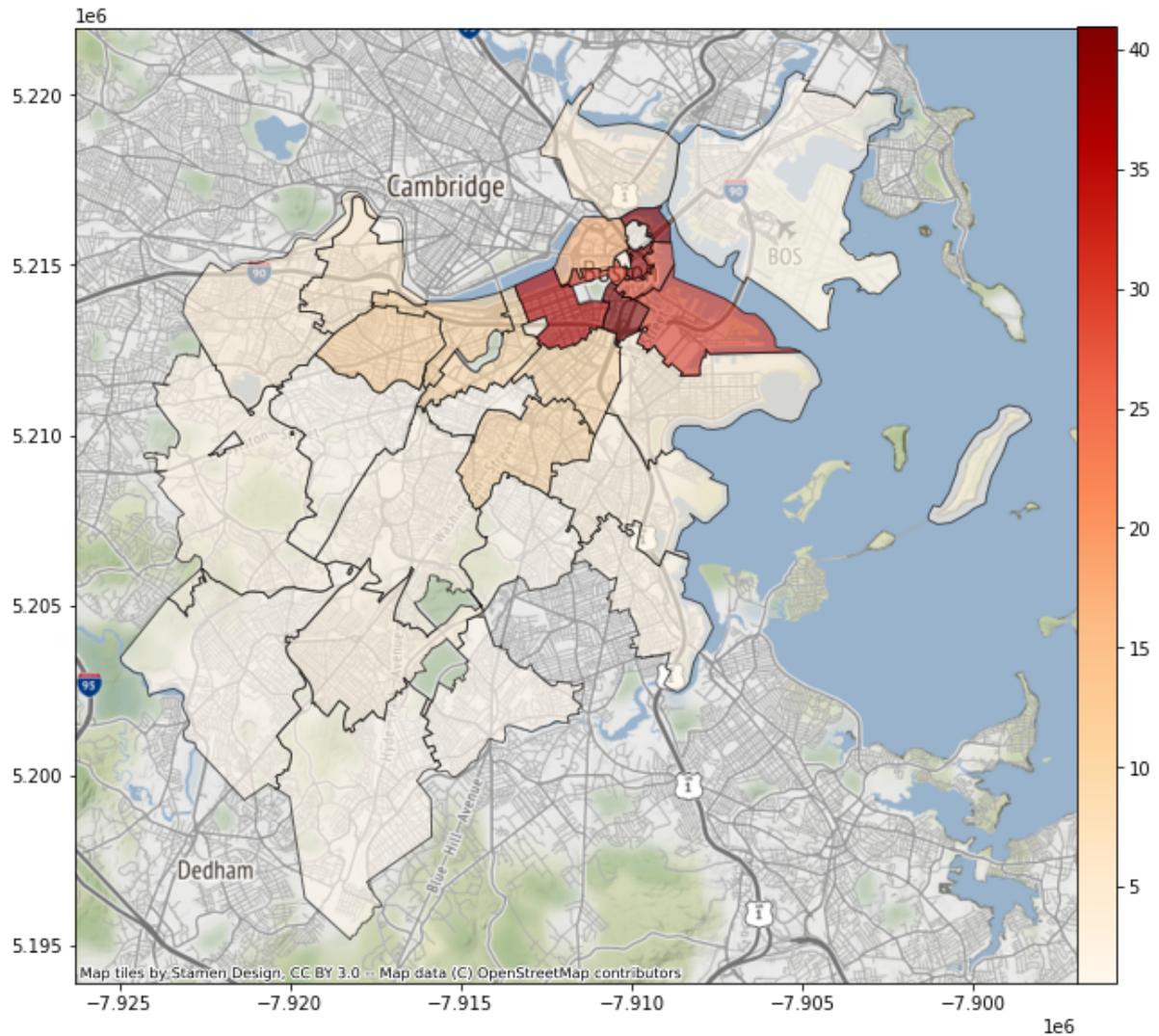

Figure 2. Map indicating concentration of business activity, intensity indicated by color and the significance level by size (darker color indicates higher number of concentrated businesses).

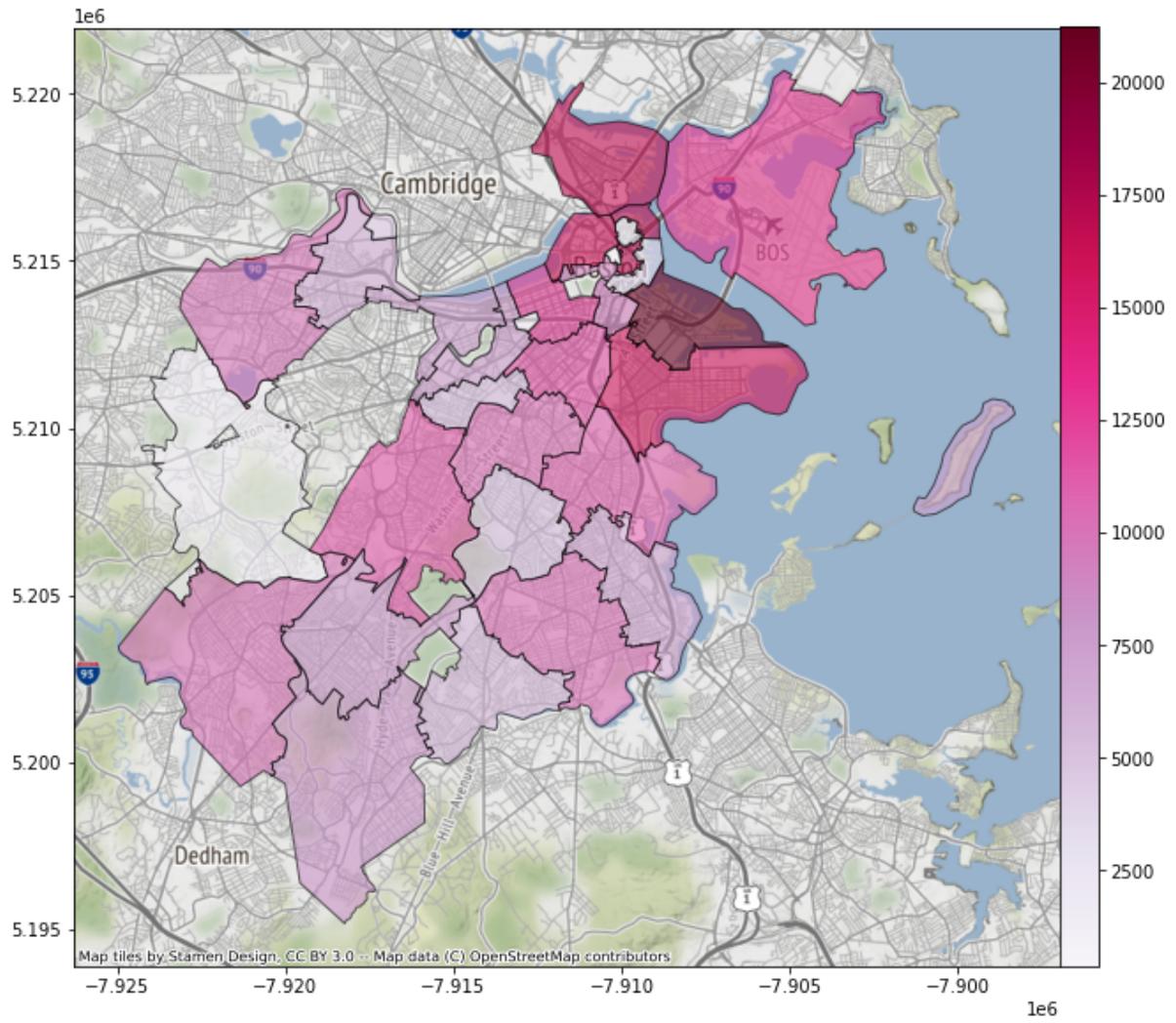

Figure 3. Map indicating location of vacant units in the municipality of boston; darker color indicates higher number of vacancies.

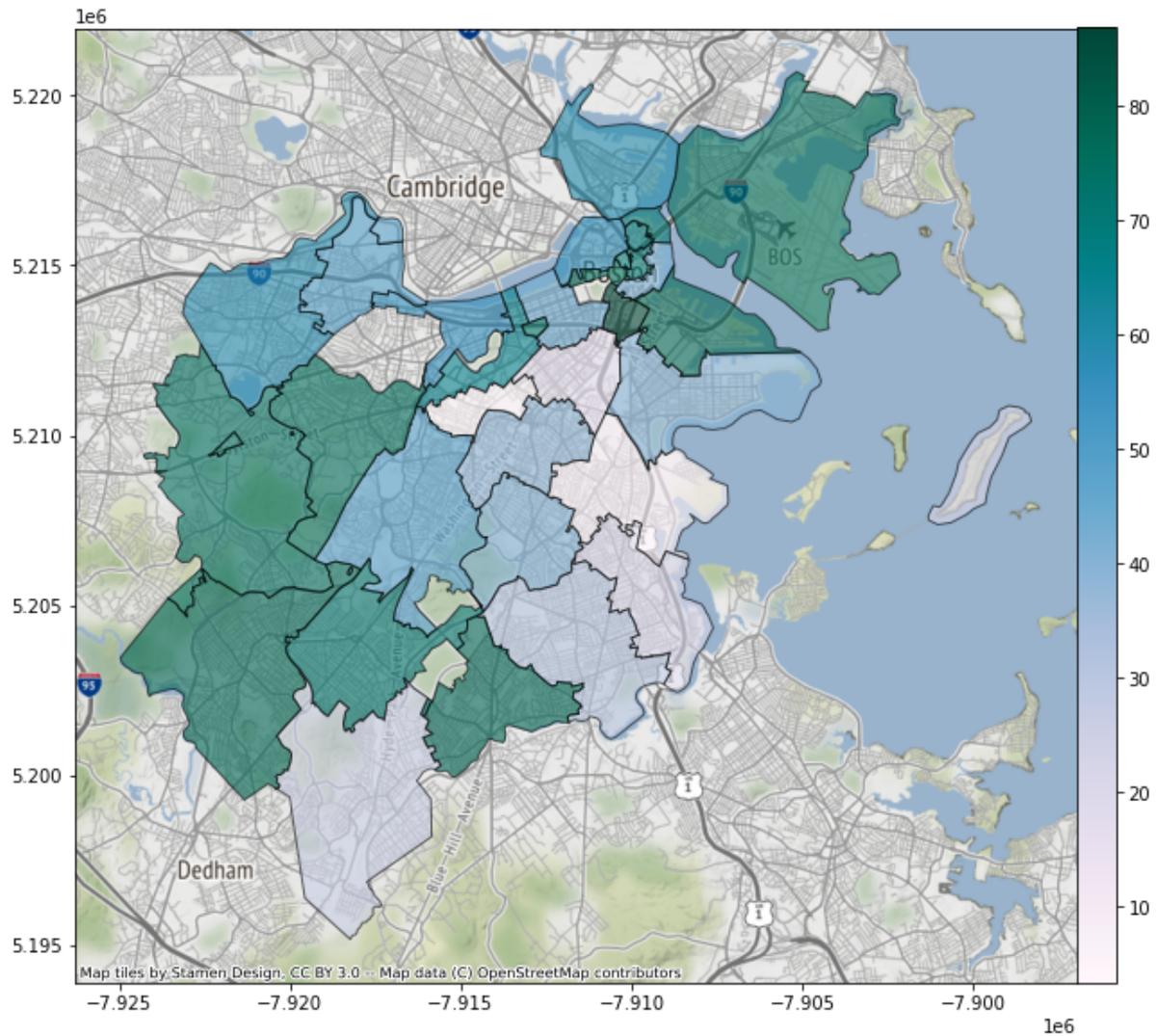

Figure 3. Map indicating White population as a percentage of total populations.

## 4. Conclusions and Next Steps

First, we analyzed the data in the spatial context to perform an illustration of innovation activity mapping for policymakers of Boston city. We observed clustering phenomena by downtown's and adjacent neighborhoods' tendency on higher innovation activity among 23 Bostonian neighborhoods. Exercising innovation heat maps, based on data coming from different sources, helped us cross check the communities that outperform and concentrate the highest number and ratings of businesses, highest number of investment in business permits etc. As discussed in Section 3.2 we found that Downtown, Wharf District, North End, Back Bay, the top performing zip codes in terms of innovation, are areas where the most civic buildings are located (see Fig. 1).

Second, high innovation/entrepreneurial activity can contribute to more dynamic and appealing communities that can result in greater property values. By equally supporting all neighborhoods to invest in the development of new mixed-use, business projects, the renovation and beautification of existing ones, the city of Boston could mitigate the disparities among districts in the same city. Despite the fact that in the past, the city has supported several initiatives to encourage and spread innovation and technology within low-income neighborhoods such as the 'Dudley Square-Upham's Corner Corridor', a vibrant zone within the Roxbury neighborhood, as well as the Roxbury Innovation Center (RIC), these neighborhoods don't appear to have the same dynamic as the downtown located areas.

Third, the maps are created to call for more attention on structurally weak neighborhoods by policymakers and the communities to think again. However, for the next steps of this research, we suggest a comparison analysis approach, which would use the Great Boston metropolitan area and compare it with another area of similar scale. This would allow us to provide more meaningful insights to local authorities. We would also suggest a normalization of the business permit data, using a measure of "x USD per business" spent on business-related projects. In this way, we could capture neighborhoods' size, population, and business density patterns to find a uniform measure in comparing them.

Fourth, we drew districts' socio-economic diagrams capturing vacant units, concentration of White, Black and Hispanic population as a percentage of total populations etc. We believe there is more room for investigating community-level and opportunities at a larger scale, the metropolitan area of Boston due to the diverse populations in terms of ethnicity and race across the different cities. An analysis on the initial features across Boston's neighborhoods revealed significant opportunities for identifying innovation determinants. A question that remains unanswered is what other variables could capture the innovation activity.

Looking only at the white population ratio, vacancy and business concentration mapping is a starting point for our study. In a future study, it would be worth it focusing more on different ethnicities and races, land use and amenity locations. A following study should consider incorporating data extracted by social media activity, such as twitter, foursquare etc., mobility data but also data regarding internet and energy use.